\documentclass[aps,prb,reprint,superscriptaddress]{revtex4-2}
\usepackage{amsmath}
\usepackage{graphicx}
\usepackage{bm}

\usepackage{color}
  
\usepackage{lineno}

\begin{document}


\title{The role of intraband dynamics in the generation of circularly polarized high harmonics from solids}

\author{N. Klemke}
\email[]{nicolai.klemke@desy.de}

\affiliation{Center for Free-Electron Laser Science CFEL, Deutsches Elektronen-Synchrotron DESY, Notkestra\ss e 85, 22607 Hamburg, Germany}
\affiliation{Physics Department, University of Hamburg, Luruper Chaussee 149, 22761 Hamburg, Germany}

\author{N. Tancogne-Dejean}
\email[]{nicolas.tancogne-dejean@mpsd.mpg.de}
\affiliation{Center for Free-Electron Laser Science CFEL, Deutsches Elektronen-Synchrotron DESY, Notkestra\ss e 85, 22607 Hamburg, Germany}
\affiliation{Max Planck Institute for the Structure and Dynamics of Matter, Luruper Chaussee 149, 22761 Hamburg, Germany}

\author{A. Rubio}
\affiliation{Center for Free-Electron Laser Science CFEL, Deutsches Elektronen-Synchrotron DESY, Notkestra\ss e 85, 22607 Hamburg, Germany}
\affiliation{Max Planck Institute for the Structure and Dynamics of Matter, Luruper Chaussee 149, 22761 Hamburg, Germany}

\author{F. X. K\"artner}
\affiliation{Center for Free-Electron Laser Science CFEL, Deutsches Elektronen-Synchrotron DESY, Notkestra\ss e 85, 22607 Hamburg, Germany}
\affiliation{Physics Department, University of Hamburg, Luruper Chaussee 149, 22761 Hamburg, Germany}
\affiliation{The Hamburg Centre for Ultrafast Imaging, Luruper Chaussee 149, 22761 Hamburg, Germany}

\author{O. D. M\"ucke}
\affiliation{Center for Free-Electron Laser Science CFEL, Deutsches Elektronen-Synchrotron DESY, Notkestra\ss e 85, 22607 Hamburg, Germany}

\date{\today}

\begin{abstract}
Recent studies have demonstrated that the polarization states of high harmonics from solids can differ from those of the driving pulses. To gain insights on the microscopic origin of this behavior, we perform one-particle intraband-only calculations and reproduce some of the most striking observations. For instance, our calculations yield circularly polarized harmonics from elliptically polarized pulses that sensitively depend on the driving conditions. Furthermore, we perform experiments on ZnS and find partly similar characteristics as reported from silicon. Comparison to our intraband-only calculations shows reasonable qualitative agreement for a below-band-gap harmonic. We show that intraband dynamics predict depolarization effects for higher field strengths. For harmonics above the band gap, interband dynamics become important and the high-harmonic response to elliptical excitation looks systematically different. Our work proposes a method to distinguish between different high-harmonic generation mechanisms and it could pave the way to compact solid-state high-harmonic sources with controllable polarization states.
\end{abstract}

\maketitle

\section{Introduction}
\raggedbottom
\noindent High-harmonic generation (HHG) is a highly nonlinear optical process in which many photons of an ultrashort laser pulse are upconverted to one photon of much higher energy. In atomic gases, where it has been first discovered \cite{Mcpherson1987}, this process is well described by a three-step model which takes into account ionization, subsequent acceleration of the free electron in the laser field and recombination of the electron with its parent ion. In the last step, the acquired energy of the electron is emitted as a highly energetic photon \cite{Krausz09}. 
The harmonic yield decreases strongly with elliptical driver polarization, which was early understood as an indication for the validity of the three-step model, because it precisely predicts the reduction of probability for the free electron wave-packet to return to its parent ion \cite{Dietrich94,Antoine97}. The harmonic yield vanishes with circularly polarized excitation and the generation of circularly polarized harmonics with other methods has evolved to a lively topic. Elaborate schemes for circularly polarized HHG have been presented, for instance HHG with counter-rotating circularly polarized bi-color pulses \cite{Eichmann1995,Fleischer2014}, with non-collinear counter-rotating circularly polarized pulses \cite{Hickstein2015} and the combination of two orthogonal linearly polarized HHG beams with an appropriate phase shift \cite{Azoury2019}. 

In crystals \cite{Ghimire11,Ghimire2019} the simple recollision-physics picture of gas HHG does not hold. Here, electrons are never really free and their energy dispersion is given by the band structure of the solid target, that consists of many bands with different momentum-dependent curvatures and probabilities of transitions between the bands. Moreover, the Coulomb potential cannot be neglected, as the electrons travel in matter, and the single-active electron approximation needs to be replaced by the assumption that electrons are independent particles, which is not always a good approximation, for instance in so-called strongly-correlated materials \cite{TD2018}. 

The deviations of the band dispersion from the quadratic free-electron dispersion cause the electrons to move in a nonlinear fashion, thereby emitting higher frequency components than the fundamental driving field contains. This is a new type of HHG mechanism that cannot be found in HHG from gaseous atoms and is called 'intraband' mechanism. In contrast, the 'interband' mechanism describes the radiation emitted upon transition from one band to another and is somewhat more similar to the emission of higher harmonics during the recombination step in gaseous atoms \cite{Vampa2017,Ghimire2019}. 

The response of solid-state HHG to elliptically polarized driving fields has been found to strongly differ from the atomic case. For instance, it has been demonstrated experimentally in MgO and graphene, that the harmonic yield could be enhanced when changing from linear to elliptical driving polarization \cite{You2017,Yoshikawa2017}. Subsequent works studied the polarization states of the emitted harmonics and found that circularly polarized harmonics can be generated from circularly \cite{NicolasNC17,klemke19,Saito17} and elliptically \cite{NicolasNC17,klemke19} polarized single-color driving pulses. In the first case, the harmonics' polarization states can be understood by a group-theoretical analysis leading to selection rules for each of the crystallographic groups, which was derived already 50 years ago \cite{Tang71}. The polarization states of the emitted harmonics driven by elliptically polarized fields, however, were shown to depend sensitively on the driving ellipticity and crystal rotation \cite{klemke19,NicolasNC17}. Moreover, they were intensity dependent and therefore directly dependent on the precise carrier dynamics \cite{klemke19,NicolasNC17}. All this is well reproduced with a time-dependent density functional theory (TDDFT) description which includes the full band and crystal structure and does not require any \textit{a-priori} assumptions to match the experiments \cite{klemke19}.  On the other hand, these calculations are costly and it is not always straightforward to extract an intuitive physical picture from these complex simulations. 

In this work, we therefore attempt to isolate a single mechanism and study its consequences on the harmonics' polarization states. Our aim is that this reduction to a simplified physical model will allow us to qualitatively understand the microscopic origin of some of the observed phenomena and that this helps interpreting the obtained experimental results.

While both intra- and interband mechanisms are intrinsically coupled \cite{Golde2008}, it is understood that the interband mechanism only contributes for photon energies above the band gap and for a reasonably high joint density of states \cite{NicolasPRL17}. Because circularly polarized harmonics from elliptically polarized driving fields have also been demonstrated below the band gap and with a low joint density of states \cite{klemke19}, we will focus our attention here on the intraband mechanism. Intraband-only calculations have been successfully utilized to reproduce the linear relationship of the cutoff energy to the driving field \cite{Feise1999,Ghimire11,Mucke11}, the six-fold rotational symmetry of HHG spectra in the three-fold symmetric crystal GaSe \cite{Kaneshima18}, anisotropic HHG emission in ZnSe \cite{Lanin2019} as well as to reconstruct the band structure \cite{Luu15,lanin2017mapping} and the Berry curvature \cite{Luu18} of SiO$_2$. However, in solid-state HHG, simplified models that consider only intraband dynamics have so far not been used to study the effects of elliptical polarization. It is clear that such a model ignores influences from dephasing \cite{Vampa14}, wave packet spreading \cite{NicolasNC17}, HHG from multiple bands \cite{Hawkins15}, contribution from holes \cite{Lanin2019}, along with effects of the subcycle ionization dynamics \cite{Hawkins13}.

We want to emphasize that the intraband mechanism was also used to model high-harmonic emission with the highest photon energies reported from solids to date ($\approx$ 40\,eV) \cite{Luu15}. Indeed, there are transparency regions in solids, for which the joint density of states (JDOS) goes to zero and interband recombination is not allowed as no pair of valence and conduction band exists with such energy. In these regions, only intraband harmonics can appear, and our results should also apply there. So, even if we only discuss low-order harmonics in this work (to compare with our experimental data), our findings should be applicable to any intraband-only generation of higher energy photons, potentially enabling circularly polarized harmonics up into the extreme ultraviolet (XUV) spectral region.

The rest of the paper is organized as follows. Having reviewed the theoretical and experimental methods in the next section, we present simulation results from a simple tight-binding-type band structure in Sec. III. In Sec. IV we present measurements on ZnS and compare these measurements to our calculations. Finally, we summarize the work and draw our conclusions in Sec. V.

\section{Methods}
\subsection{Theoretical model}
\noindent Here we start by considering the dynamics of an electron wave packet in a single band. The current density $\textbf{j}$ at time $t$ can be described as

\begin{align}
\textbf{j}(t) = -\int_{\mathrm{BZ}} e \textbf{v}_\textbf{k}(t) n_\textbf{k}(t) \text{d\textbf{k}} . \label{eq:j1}
\end{align}
Here, $\mathrm{BZ}$ refers to the first Brillouin zone, $e$ is the electron charge,   $n_\textbf{k}$ is the charge distribution in $\textbf{k}$-space and $\textbf{v}_\textbf{k}$ is the $\textbf{k}$-dependent electron velocity. The latter consists of two terms, one of which is coined the anomalous velocity that contains the Berry curvature \cite{Xiao2010,Liu2017}. For the square lattice in Sec. III, the Berry curvature is zero because of symmetries and for ZnS in Sec. IV, we have confirmed that the influence of a band-averaged Berry curvature on the studied odd harmonic is neglible. Too keep our discussion as simple as possible, we will therefore neglect the anomalous velocity term in this paper, which is in agreement with other recent works that utilized this model to study odd orders in solid HHG \cite{Kaneshima18,lanin2017mapping,Lanin2019,Luu15}.  

Assuming a fully localized electron wave packet at $\textbf{k}(t)$, i.e. $n_\textbf{k}(t) = \delta(\textbf{k}-\textbf{k}(t))$, and inserting the definition of the electron velocity (without the anomalous velocity term) $\textbf{v}_\textbf{k}= \frac{1}{\hbar}\frac{\text{d}\mathcal{E}_\textbf{k}}{\text{d}\textbf{k}}$, Eq. (\ref{eq:j1}) simplifies to

\begin{align}
\textbf{j}(t) = -\frac{e}{\hbar}\frac{\text{d}\mathcal{E}_\textbf{k}}{\text{d}\textbf{k}}\bigg\rvert_{\textbf{k}=\textbf{k}(t)}.
\end{align}

\noindent $\mathcal{E}_\textbf{k}$ is the conduction band dispersion. Under these assumptions, the emitted electric field $\textbf{E}^{\text{HH}}(t)$ originating from an intraband current is

\begin{align}
\textbf{E}^{\text{HH}}(t) \propto \frac{\text{d}\textbf{j}(t)}{\text{d}t} 
&= - \frac{e}{\hbar}\frac{\text{d}^2\mathcal{E}	_\textbf{k}}{\text{d}\textbf{k}^2}\frac{\text{d}\textbf{k}}{\text{d}t}\bigg\rvert_{\textbf{k}=\textbf{k}(t)} \nonumber\\ 
&= e^2\left(\frac{1}{m_\textbf{k}^{\ast}}\right)\bigg\rvert_{\textbf{k}=\textbf{k}(t)} \cdot \textbf{E}_\text{L}(t). \label{eq:final1band}
\end{align}

\noindent $\textbf{E}_\text{L}(t)$ denotes the driving laser field.
Furthermore, $m_\textbf{k}^{\ast}$ is the effective mass tensor. Here we have used the acceleration theorem $\textbf{k}(t) = -\frac{e}{\hbar} \int_{-\infty}^{t} \textbf{E}_\mathrm{L}(t')dt'$ and with it the assumption that the electron is initially located at the $\Gamma$-point.

\noindent Finally, the emitted high-harmonic spectrum can be calculated as 

\begin{align}
\text{I}_\text{HH}(\omega) \propto \left|\text{FT}[\textbf{E}_{\text{HH}}(t)]\right|^2.
\end{align}

\noindent It can be seen from Eq. (\ref{eq:final1band}) that the nonlinear evolution of $\frac{\textrm{d}^2 \mathcal{E}_\textbf{k}}{\textrm{d} k^2}$ is the source for non-perturbative emission of higher frequency content. When happening in repetition over multiple laser cycles, this emission consists of high harmonics of the driving laser frequency \cite{Golde2008,Ghimire11}.

In the following discussion we will not discuss $E_x$ and $E_y$ but convert them into the parallel and perpendicular field components defined with respect to the major axis of the polarization ellipse of the driving field, i.e., $E_\parallel$ and $E_\bot$. Whenever we discuss harmonic order $n$ or the corresponding electric field $\textbf{E}_n$, we have band pass filtered the Fourier transform of the total electric field in a window of $nf \pm 0.3\mathrm{f}$ when f is the center frequency of the driving field and $n$ is the harmonic order. By finding the axes of minimum ($\alpha_{min}$) and maximum ($\alpha_{max}$) harmonic yield $I_n$, we calculate the harmonic ellipticity as

\begin{align}
|\epsilon_{n}| = \sqrt{\frac{I_n(\alpha_{min})}{I_n(\alpha_{max})}}.
\end{align}
This approach resembles the experimental method to rotate a polarizer in order to determine $|\epsilon_n|$. In our simulations, we use the driving field
\begin{align}
\textbf{E}_\textrm{L}(t) = \frac{\tilde{E}(t)}{\sqrt{1+\epsilon^2}} \begin{pmatrix}
 \cos(\omega t)\\
\epsilon\sin(\omega t)
\end{pmatrix},
\end{align}
where $\tilde{E}(t)$ is a Gaussian envelope with a FWHM pulse duration of 70\,fs and a central wavelength of 2100\,nm. The field is rotated by an angle $\theta$  by multiplying $\textbf{E}_\textrm{L}$ with the rotation matrix. Throughout this paper we keep the driving field strength below the threshold above which Bloch oscillations appear. This is compatible with experimental conditions for this driving wavelength \cite{klemke19}. 

DFT calculations for bulk ZnS were performed with the Octopus code \cite{Marques2003octopus,Castro2006octopus,Andrade2015real,Tancogne2020}, using a lattice parameter of 5.41\,\AA, norm-conserving pseudopotentials, and a real-space grid spacing of 0.25 Bohr. We used a sampling of 21x21x21 $\mathbf{k}$-points to sample the Brillouin zone and we approximated the exchange-correlation term using the functional proposed by Tran and Blaha  \cite{Tran2009}.

\subsection{Experiment}
\noindent In Section IV we also present experimental results on bulk ZnS. Those are done with a Ti:Sapphire-pumped OPA source that produces CEP-stable, 70\,fs pulses with a wavelength of 2100\,nm. We use a peak electric field strength of approximately 1\,V/nm in matter. Both experimental setup and procedure are the same as described extensively in Ref. [\onlinecite{klemke19}] and its supplement. Importantly, we measure the harmonics' ellipticity by inserting a Rochon polarizer between sample and spectrometer and record the spectra for different rotations of the polarizer. The ellipticity is then calculated by fitting the harmonic yield over polarizer rotation with a $\sin^2$-function and calculating $|\epsilon_{n}|=\sqrt{I_{min}/I_{max}}$, with $I_{max}$ and $I_{min}$ being the minimum and maximum intensities of the harmonic over polarizer rotation. This is only an upper limit to the ellipticity, as will be discussed in Section IV. To evaluate the degree of polarization of the harmonics, we use a Fresnel rhomb and insert it between sample and Rochon polarizer. 

\section{Square lattice}

\noindent First we will discuss a two-dimensional tight-binding band structure  
\begin{align}
\mathcal{E}_\mathbf{k} = \frac{\hbar^2}{4 a^2 m_e} \left[ 1+ \sum_m c_m [cos(m k_x a) + cos(m k_y a)] \right]. \label{eq:TBband}
\end{align}
We set all $c_m$ zero except for $c_1 = -0.95$ and $c_3 = -0.05$. These coefficients have previously been used in the one-dimensional case to theoretically model HHG from ZnO with intraband dynamics alone \cite{Ghimire11}. We use a lattice constant of a = 5.4\,\AA{}. 

Equation (\ref{eq:TBband}) describes a square lattice. We call the axes parallel to $k_x$ and $k_y$ $\Gamma$X and the ones rotated by $45^{\circ}$ $\Gamma$K. The peak electric field is 2\,V/nm. Figure 1(a) shows the calculated high-harmonic spectra with linear polarization along $\Gamma$X and $\Gamma$K. With this band structure and our driving conditions, the harmonic signal is maximized along $\Gamma$X, where harmonics are generated up to the 11$^{\mathrm{th}}$-order (HH11). Along $\Gamma$K, the overall harmonic yield is lower and the also the cutoff is reduced. Figure 1(b) (bottom panel) depicts the perpendicular and parallel components of HH5 and HH9 versus crystal rotation angle. The signal is four-fold symmetric, as expected from a cubic structure. For polarization along the symmetry axes, the emitted field is completely parallel to the driving field. However, for sample rotations between 0$^{\circ}$ and 45$^{\circ}$, the emitted harmonic fields contain a small perpendicular component. This is a consequence of different band curvatures along $x$- and $y$- components of the driving field (Eq. (\ref{eq:final1band})). Because in this case the respective relative phases $\varphi$ (top panel of Fig. 1(b)) are close to 0 for all sample rotations, the emitted fields stay linearly polarized (center panel of Fig. 1(b)) but are rotated with respect to the driving field. A behavior similar to this has already been observed experimentally \cite{klemke19,You19}.

\begin{figure}[tb]
\includegraphics[width=\columnwidth]{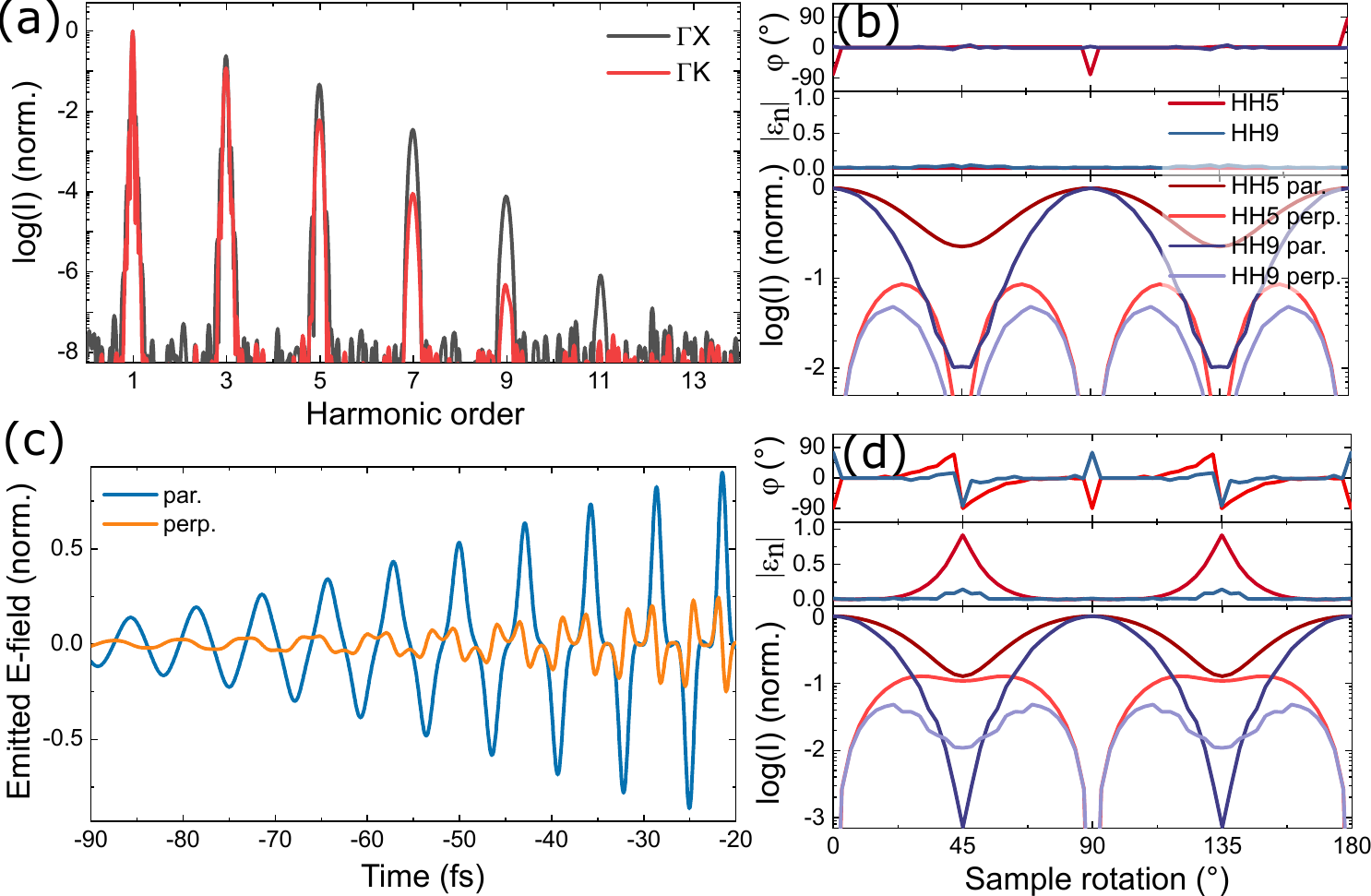}
\caption{(a) Spectra along the major symmetry axes $\Gamma$X and $\Gamma$K ($\epsilon =0$). (b) top panel: relative phase $\varphi$ between $E_\parallel$ and $E_\bot$ of HH5 and HH9; center panel: Respective harmonic ellipticities $\epsilon$; bottom panel: Individual intensity components of HH5 and HH9 parallel and perpendicularly polarized to the driving field for different sample rotations. ($\epsilon = 0$) (c) Excerpt of the emitted electric field components parallel and perpendicular to the driving major axis along $\Gamma$K. ($\epsilon =0.15$) (d) same as (b) but with $\epsilon = 0.15$.}
\end{figure}

Figure 1(c) shows the temporal evolution of $E_\parallel$ and $E_\bot$ with a driver ellipticity $\epsilon = 0.15$ and major axis along $\Gamma$K. Only an excerpt of the rising edge of the pulse is shown - the highest field strength is reached at 0\,fs. Although the perpendicular component is much weaker due to the low ellipticity of the driving field, it shows non-sinusoidal behavior starting at around -70\,fs which is earlier than the parallel component, where clear non-sinusoidal components start to arise only at around -50\,fs. This illustrates how the high harmonics can have totally different polarization states than the driving field.

The harmonics' behavior for the same driving ellipticity $\epsilon = 0.15$ as function of the crystal rotation angle is depicted in Fig. 1(d). Now the relative phases between $E_\parallel$ and $E_\bot$ evolve in a more complicated way, reaching $90^{\circ}$ along $\Gamma$K. Hence, the ellipticities (center panel of Fig. 1(d)) peak for this sample rotation. HH5 becomes circularly polarized while HH9 exhibits only little ellipticity. For excitation away from a major symmetry axis, the fields are again rotated with respect to the driving laser. In summary, the harmonics' polarization states differ among individual harmonics and sensitively depend on the crystal rotation and the driving ellipticity. Importantly, our results show that pure intraband dynamics are sufficient to produce circularly polarized harmonics from elliptically polarized driving pulses, as we have observed experimentally from Si \cite{klemke19}.

\begin{figure}[tb]
\includegraphics[width=\columnwidth]{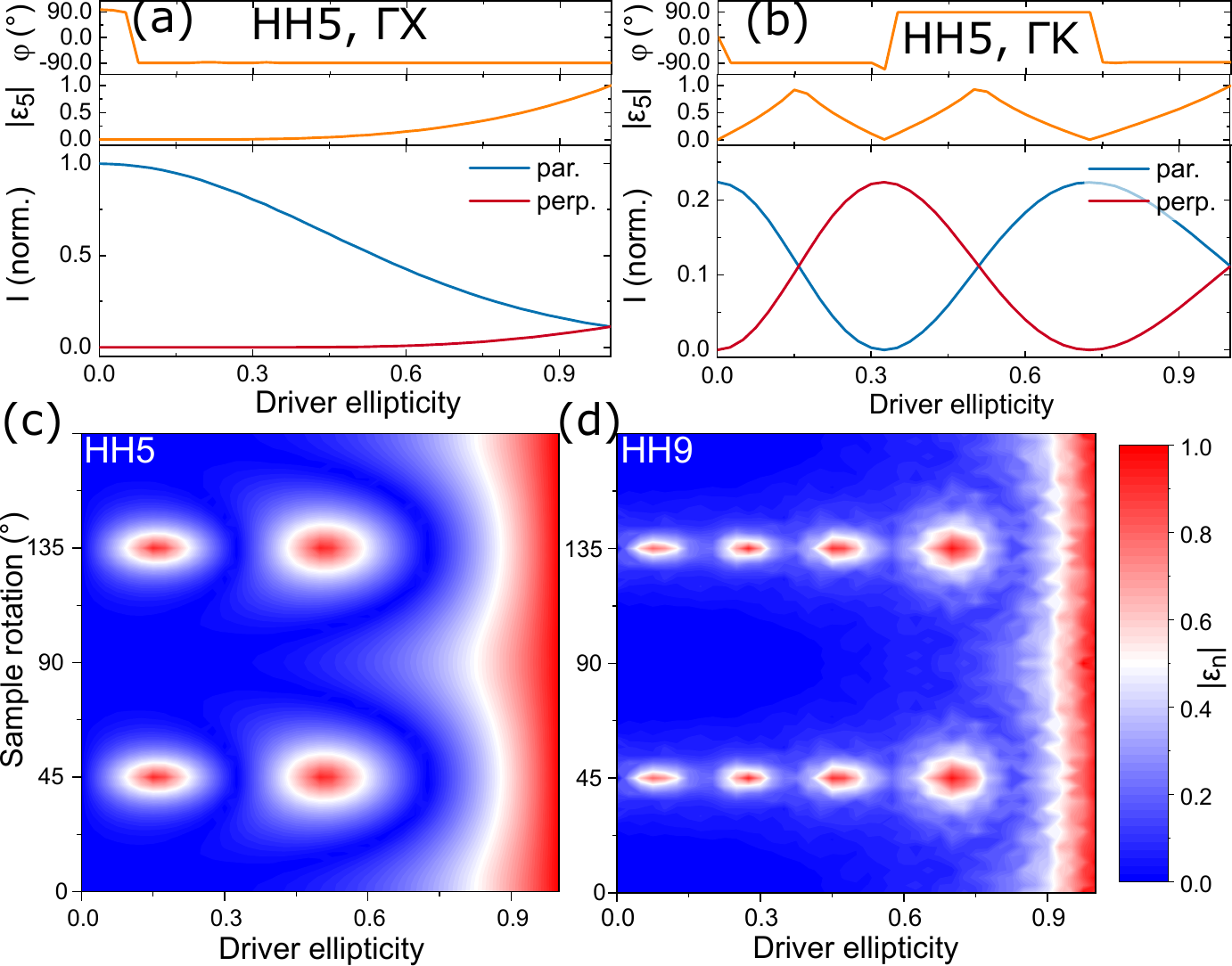}
\caption{(a)\&(b): Ellipticity dependent relative phases between $E_\bot$ and $E_\parallel$   (top panel), corresponding ellipticities (center panel) and intensities of parallel and perpendicular components of HH5 along $\Gamma$X (a) and $\Gamma$K (b) (bottom panel). Ellipticities of HH5 (c) and HH9 (d) in dependence of sample rotation and driver ellipticity.0$^{\circ}$ and 90$^{\circ}$ refers to $\Gamma$X, 45$^{\circ}$ and 135$^{\circ}$ to $\Gamma$K.}
\end{figure}

Next, we extend the analysis of the harmonics' polarization states by varying the driving ellipticity further. Figs. 2(a) and (b) show the evolution of HH5's polarization versus driving ellipticity along $\Gamma$X (a) and $\Gamma$K (b). As discussed above, $\Gamma$X is the direction to most efficiently generate harmonics. To drive the electrons away from that axis with an introduced ellipticity means - at least for this simple band structure - to generate harmonics less efficiently. The total harmonic yield therefore decreases (bottom panel of Fig. 2(a)) with very little rise of the perpendicular component. Hence, although the relative phase (top panel) is $90^{\circ}$ for most driver ellipticities, the harmonic ellipticity is low. For circular excitation HH5 becomes also circular. In fact, all harmonics become circular for circular excitation, and we even find that our simple model predicts that subsequent harmonics have alternating helicities - which is required by selection rules for cubic materials \cite{Tang71} and was recently confirmed experimentally \cite{Saito17,klemke19}.

When the driving major axis is set along $\Gamma$K (Fig. 2(b)), harmonics are generated least efficiently. As a consequence, the perpendicular component increases strongly with small ellipticity values (bottom panel of Fig. 2(b)). At $\epsilon = 0.17$, $E_\bot$ and $E_\parallel$ have the same magnitude and - because $\left|\varphi\right|$ is 90$^{\circ}$ - HH5 becomes circularly polarized (center panel). Further increase of the driving ellipticity causes $E_\bot$ to dominate, which rotates the harmonics' major axis by 90$^{\circ}$. When the perpendicular component peaks, conditions are reversed and the parallel component rises again. Note that the relative phase is flipped in this case, which reverses the helicity of the harmonics' polarization ellipse. The ellipticity of HH5 peaks again at $\epsilon = 0.55$ and $E_\parallel$ dominates for even more elliptically polarized drivers. At circularly polarized excitation, HH5 becomes circular once again. Qualitatively, we have observed very similar behavior to this in Si (compare Figs. 4 and S10 in Ref. [\onlinecite{klemke19}]). 

For polarization along this axis $\Gamma$K, the total yield of HH5 stays constant for all driver ellipticities. We would like to emphasize that - contrary to HHG from atomic gases - the intraband mechanism does not necessarily predict the harmonic yield to decrease with increasing ellipticity. The harmonic yield is a sole consequence of the band curvatures at different $\textbf{\text{k}}$-values which can even be higher for elliptical polarization. Similar harmonic yields for linear and circular excitation have been reported in SiO$_2$ \cite{klemke19} and GaSe \cite{Saito17} with mid-infrared excitation. Interband transitions however, will be reduced with elliptically polarized fields because the peak electric field is reduced by $E_{\textrm{elli}}/E_{\textrm{lin}} = \sqrt{1/(1+\epsilon^2)}$. Therefore, less conduction-band electrons should contribute to the HHG current. But even here the effect on the harmonic yield is unclear, because less electrons could also cause less dephasing to happen and thereby increase the harmonic yield. All this should be dependent on the exact driving conditions and the band structure and is not exactly understood at this point.

Figures 2(c) and (d) depict the full polarization maps of HH5 and HH9 versus driving ellipticity and sample rotation. The previous discussion is summarized in these plots. The conditions of high $|\epsilon_n|$ are seen as 'islands' along $\Gamma$K. For higher harmonics, these islands are more sensitive to the exact driving conditions and more islands appear. Qualitatively, this is  what has been observed in Ref. [\onlinecite{klemke19}]. For other sample rotations, the harmonics stay largely linearly polarized due to a small relative phase between $E_\bot$ and $E_\parallel$. However, they are often rotated with respect to the driving field (not shown). As required for a cubic system, all harmonics become circularly polarized for $\epsilon=1$ independent on the sample rotation.

\section{Zinc sulfide}\label{sec:ZnS}
\noindent After having studied a simple model band structure, we can address a real material and compare our simulation results with measurements. We have chosen to investigate 50-$\mu$m-thin, (100)-cut ZnS, because its lower conduction band is well isolated from the others and therefore our one-band model could constitute a reasonable approximation. Since harmonics below the band gap should be produced predominantly by intraband dynamics \cite{NicolasPRL17}, we will focus our attention in the following on HH5. For our peak electric field strength, the harmonics are generated non-perturbatively, as we have confirmed by studying harmonics' yield versus driving field strength \cite{Ghimire11}. In the simulations, unless otherwise noted, we use the same peak electric field strength as in the experiment. The band structure of ZnS has been constructed as described in Sec. II.A.

\begin{figure}[tb]
\includegraphics[width=\columnwidth]{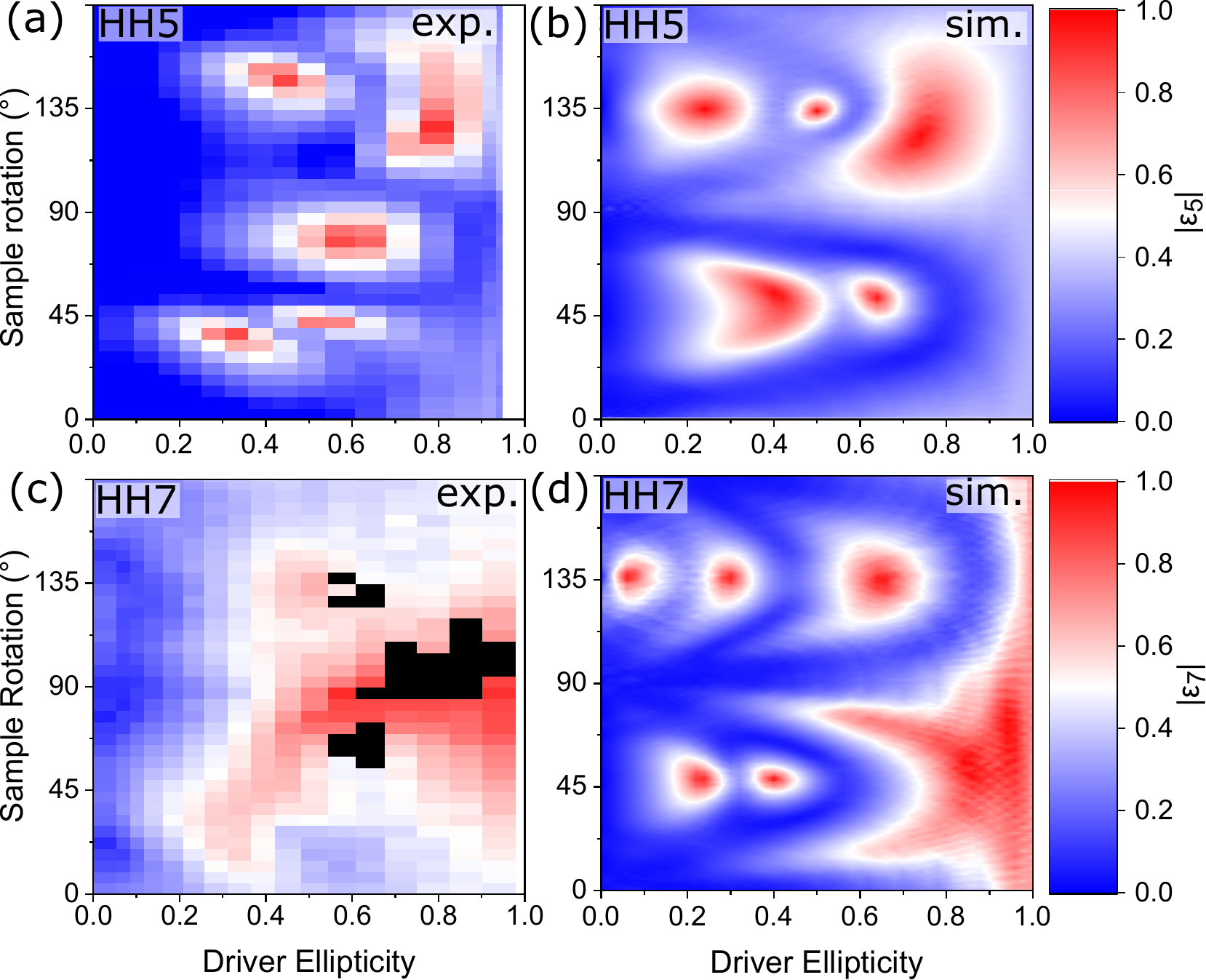}
\caption{Full experimental (a \& c) and theoretical (b \& d) results of the ellipticity of HH5 and HH7 in dependence of driving ellipticity and sample rotation from ZnS. Datapoints in which the signal to noise ratio is lower than 2.57 (99\% confidence interval) are marked in black.}
\end{figure}

ZnS has a zinc-blende crystal structure and hence, is not inversion symmetric. In the experimental harmonic spectra, the lack of inversion symmetry manifests itself in the generation of even-order harmonics. Because the single band model employed in this work cannot produce even harmonics \cite{Kaneshima18,Luu18}, we will investigate here only the odd harmonics.
 
Figure 3(a) and (b) show the measured and calculated $|\epsilon_{\textrm{5}}|$ in dependence of sample rotation and driving ellipticity. Due to the zinc-blende crystal structure, neither the experimental nor the calculated data show a four-fold symmetry. A somewhat peculiar aspect of this crystal structure is that harmonics are elliptically polarized with circular excitation. This is required by selection rules for the zinc-blende symmetry group \cite{Tang71} and is confirmed in both our experimental data and calculations. For elliptical excitation, we once again find islands of high $|\epsilon_5|$, both in the simulated as well as in the measured data. Some features of the experimental data are qualitatively well reproduced in the simulations. This is especially true for the asymmetric elongated island around $\epsilon=0.8$ and $110^{\circ} <\theta < 150^{\circ}$.  Also the two islands along $\theta \approx 45^ {\circ}$ can be found both in experiment and simulations. The experiments show a circularly polarized HH5 for $\epsilon \approx 0.6$, $\theta \approx 80^{\circ}$ which is not covered in the simulation. It can be expected that electron-electron interactions and the influence of harmonic emission due to electrons that are promoted to the conduction band at different times have a great effect on these kind of maps. Also hole dynamics can be expected to play a role \cite{Lanin2019}. Discrepancies are therefore not surprising when these effects are neglected.

For harmonics above the direct band gap, the assumption that intraband dynamics alone underlie high-harmonic emission breaks down. We can observe this in ZnS by studying HH7, which lies above the band gap. Depicted in Fig. 3(c), the measured $|\epsilon_{\textrm{7}}|$ shows a more continuous structure compared to the simulated $|\epsilon_{\textrm{7}}|$ in Fig. 3(d) and any of the ellipticity maps we have computed for intraband-only HHG. We have reported a similar qualitatively different appearance of $|\epsilon_{\textrm{7}}|$ in experimental results from silicon \cite{klemke19}. There, intraband-only harmonics (HH5 (below bandgap), HH9 (low JDOS)) showed island-type maps while HH7, produced by coupled intra- and interband dynamics, showed a more continuous structure of high harmonic ellipticity. In ZnS, the same seems to be true. Although only phenomenological at this point, this observation seems to indicate that one could discriminate between different generation mechanisms by studying the harmonics' polarization state with respect to driving pulse ellipticity.

\begin{figure}[tb]
\includegraphics[width=\columnwidth]{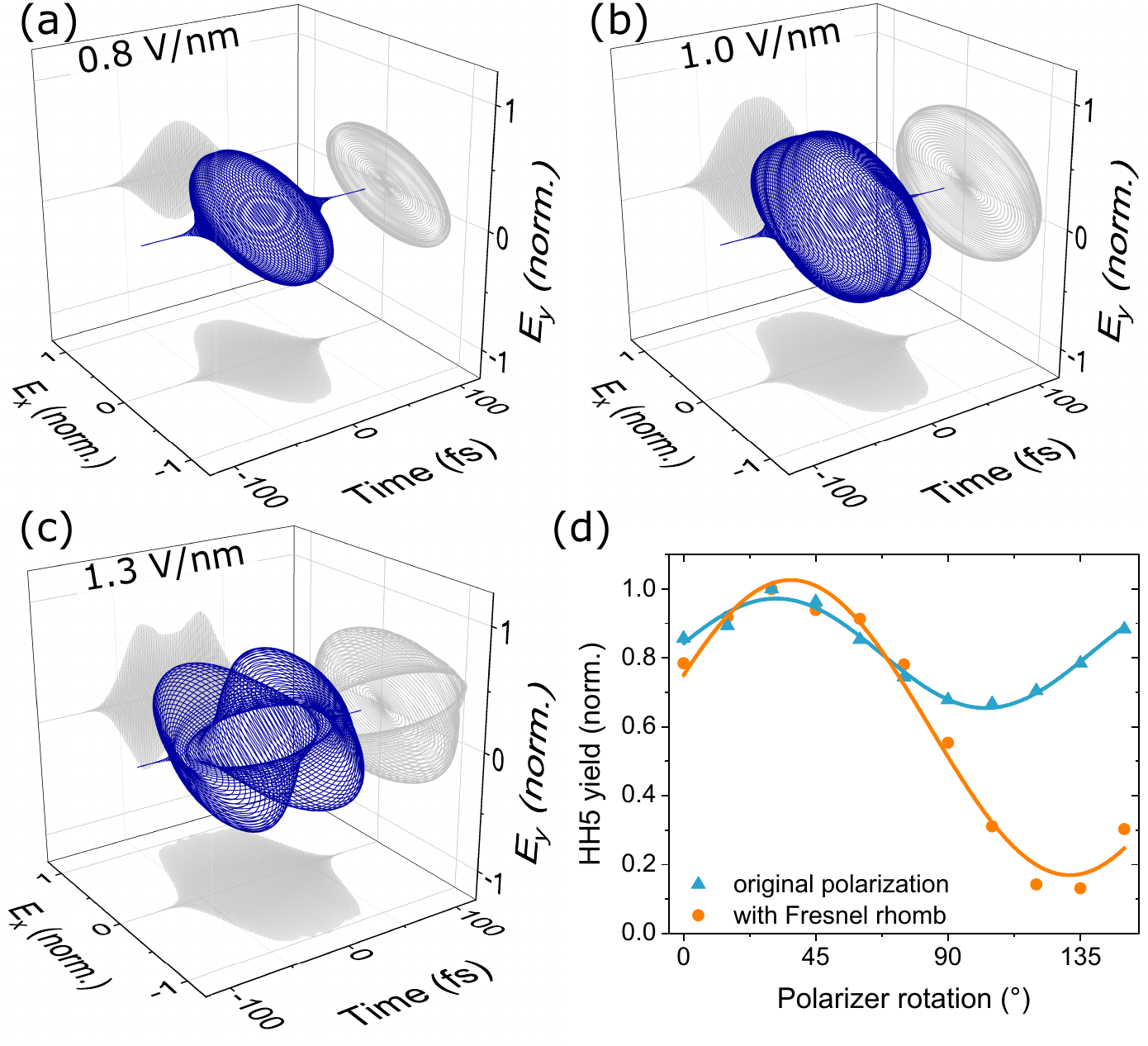}
\caption{Calculated electric fields of HH5 generated with a peak electric field strengths of 0.8\,V/nm (a), 1.0\,V/nm (b) and 1.3\,V/nm (c). Polarizer scan of HH5 with and without an additional Fresnel rhomb (d). Solid lines are $\sin^2$-fits. For all figures: $\epsilon=0.3$ and $\theta=37^{\circ}$. }
\end{figure}

Let us now comment on the ellipticity. What we have called $|\epsilon_n|$ so far is in fact an upper limit to the ellipticity. By simply rotating a polarizer and calculating $|\epsilon_n|$ from that, one would find $|\epsilon_n|=1$ for completely unpolarized light. We will show now that one can not simply assume harmonics to be fully polarized in solid HHG. Figures 4(a)-(c) depict the calculated emitted electric fields that have been band-pass-filtered around HH5 for three different driving field strengths with elliptically polarized excitation. While for the lowest field strength, the harmonic field is simply elliptically polarized, for higher field strengths the polarization dynamically evolves over the course of the pulse. This effective 'depolarization' enters naturally when the effective mass tensor impacts the polarization differently with increasing $\textbf{k}$-values in Eq. (3). For higher field strengths, the electron explores a larger region of the BZ and hence, $m_\textbf{k}^*$ can produce completely different fields. We would like to emphasize that this happens already in this simple single-particle one-band model, without any ionization effects being included.

In experiment, accessing the part of the harmonic that is circularly polarized and, with it, the degree of polarization, requires usage of an additional quarter-wave plate (see for instance Ref. \cite{klemke19}). We use a Fresnel rhomb here, which serves the same purpose. Fig. 4(d) shows two measured polarizer scans of HH5 when generated from ZnS with the same driving conditions as calculated in Fig. 4(b). The blue curve shows the unaltered HH5, exhibiting very little modulation over polarizer rotation, which yields $|\epsilon_{\textrm{5}}|= 0.84$. When a Fresnel rhomb is inserted, the same harmonic shows severe modulation, proving in this case that the original HH5 is highly polarized. Here, as predicted by the simulations in Fig. 4(b), the driving conditions are not right to drive the electrons into regions in which the varying $m_\textbf{k}^*$ changes the polarization of HH5 over the course of the pulse significantly. In any case,  depolarization effects have been discussed recently in gas HHG \cite{Barreau2018} and we predict that the influence of the band structure can cause an even stronger depolarization in solid HHG, although this remains yet to be observed. 

Note also that the original polarizer scan without the Fresnel rhomb in Fig. 4(d) reveals a slightly rotated polarization ellipse of HH5 by approx. $30^{\circ}$. The polarizer in this experiment rotates clockwise and hence, the rotation of the major axis fits well to the slightly rotated polarization ellipse in the calculation of Fig. 4(b).

\section{Conclusion}
\noindent We have demonstrated that simple intraband dynamics alone produce salient features in the polarization of HHG from solids that can also be found in experimental results. Striking are the appearances of circularly polarized harmonics with elliptically polarized excitation as well as deviations of the harmonics' major axis with respect to the driving major axes. Experimentally, after having demonstrated this behavior in cubic Si for the first time \cite{klemke19}, we have shown here that zinc-blende ZnS is another material from which one can produce circularly polarized high harmonics with elliptically polarized excitation. This suggests that this is a fundamental response of solid HHG to elliptical excitation and that it can be found in a much broader range of excitation conditions and crystals. We have discussed effects of depolarization that can result from intraband dynamics alone for high enough field strengths. We have also demonstrated differences in the polarization-state-resolved response of high harmonics above the band gap, where the interband mechanism cannot be neglected. Previously, the intraband-only model was used to successfully reproduce the XUV-spectra up to record 40\,eV photon energy from quartz \cite{Luu15}. In consequence, we predict that the island-like circularly polarized harmonics from elliptically polarized driving pulses can also be found in this spectral region. This could pave the way to relatively compact sources of circularly polarized XUV radiation. Furthermore, the direct link of intraband dynamics to solid HHG could allow for $\textbf{k}$-resolved tracking of the fastest oscillating currents that ultrafast laser pulses can generate in solids to date.

\begin{acknowledgments}
\noindent This work was supported by the European Research Council 
(ERC-2015-AdG694097), the Cluster of Excellence 'Advanced Imaging of Matter' (AIM) and SFB925 'Light induced dynamics and control of correlated quantum systems'. N. Klemke is part of the Max Planck School of Photonics supported by BMBF, Max Planck Society, and Fraunhofer Society. 
\end{acknowledgments}


\bibliography{cite}

\end{document}